\documentclass[showpacs,floatfix,citeautoscript,twocolumn,prb,superscriptaddress]{revtex4-1}

\usepackage{graphicx}
\usepackage[dvipsnames]{xcolor}
\usepackage{amsmath}
\usepackage{comment}
\usepackage{textcomp}
\usepackage{inputenc}

\begin{document}

\title{Magnetic anisotropy in single-crystalline antiferromagnetic Mn$_2$Au}

\author{Mebatsion S. Gebre}
\affiliation{Department of Materials Science and Engineering  and Materials Research Laboratory, University of Illinois at Urbana-Champaign, Urbana, Illinois 61801, United States}
\author{Rebecca K. Banner}
\affiliation{Department of Materials Science and Engineering  and Materials Research Laboratory, University of Illinois at Urbana-Champaign, Urbana, Illinois 61801, United States}
\affiliation{School of Materials Science and Engineering, Georgia Institute of Technology, Atlanta, Georgia 30332, United States}
\affiliation{Department of Chemistry, Skidmore College, Saratoga Springs, New York 12866, United States}
\author{Kisung Kang}
\affiliation{Department of Materials Science and Engineering  and Materials Research Laboratory, University of Illinois at Urbana-Champaign, Urbana, Illinois 61801, United States}
\author{Kejian Qu}
\affiliation{Department of Physics and Materials Research Laboratory, University of Illinois at Urbana-Champaign, Urbana, Illinois 61801, United States}
\author{Huibo Cao}
\affiliation{Neutron Scattering Division, Oak Ridge National Laboratory, Oak Ridge, Tennessee 37831, United States}
\author{Andr\'{e} Schleife}
\affiliation{Department of Materials Science and Engineering  and Materials Research Laboratory, University of Illinois at Urbana-Champaign, Urbana, Illinois 61801, United States}
\affiliation{National Center for Supercomputing Applications, University of Illinois at Urbana-Champaign, Urbana, Illinois 61801, United States}
\author{Daniel P. Shoemaker}
\affiliation{Department of Materials Science and Engineering  and Materials Research Laboratory, University of Illinois at Urbana-Champaign, Urbana, Illinois 61801, United States}

\begin{abstract} 

Multiple recent studies have identified the metallic antiferromagnet Mn$_2$Au to be a candidate for spintronic applications due to apparent in-plane anisotropy, preserved magnetic properties above room temperature, and current-induced Néel vector switching.
Crystal growth is complicated by the fact that Mn$_2$Au melts incongruently. 
We present a bismuth flux method to grow millimeter-scale bulk single crystals of Mn$_2$Au in order to examine the intrinsic anisotropic electrical and magnetic properties.
Flux quenching experiments reveal that the Mn$_2$Au crystals precipitate below 550$^{\circ}$\,C, about 100$^\circ$\,C below the decomposition temperature of Mn$_2$Au.
Bulk Mn$_2$Au crystals have a room-temperature resistivity of 16\,--\,19 $\mu\Omega$-cm and a residual resistivity ratio of 41. 
Mn$_2$Au crystals have a dimensionless susceptibility on the order of 10$^{-4}$ (SI units), comparable to calculated and experimental reports on powder samples. 
Single-crystal neutron diffraction confirms the in-plane magnetic structure.
The tetragonal symmetry of Mn$_2$Au constrains the $ab$-plane magnetic susceptibility to be constant, meaning that $\chi_{100}=\chi_{110}$ in the low-field limit, below any spin-flop transition. We find that three measured magnetic susceptibilities $\chi_{100}$, $\chi_{110}$, and $\chi_{001}$ are the same order of magnitude and agree with the calculated prediction, meaning the low-field susceptibility of Mn$_2$Au is quite isotropic, despite clear differences in $ab$-plane and $ac$-plane magnetocrystalline anisotropy.
Mn$_2$Au is calculated to have an extremely high in-plane spin-flop field above 30~T, which is much larger than that of another in-plane antiferromagnet Fe$_2$As (less than 1~T).
The subtle anisotropy of intrinsic susceptibilities may lead to dominating effects from shape, crystalline texture, strain, and defects in devices that attempt spin readout in Mn$_2$Au.
\end{abstract}

\maketitle

%\dps{ $[]$ for directions, $()$ for planes, $\{\}$ for family of planes, and $\langle\rangle$ for family of directions is standard.}

\section{Introduction}

% motivating AF studies
Antiferromagnets (AF) have compensated, net-zero magnetization but still host a wealth of field- and current-dependent phenomena, notably spin canting in response to an applied field \cite{moriya_anisotropic_1960}, spin flop transitions \cite{baltz_antiferromagnetic_2018, bogdanov_spin-flop_2007, carlin_field-dependent_1980}, and fundamentally higher resonance frequencies than ferromagnets \cite{siddiqui_metallic_2020, wienholdt_thz_2012, olejnik_thz_2017}. 
These THz-frequency resonances especially have attracted interest for fast data storage,\cite{rezende_introduction_2019, khymyn_antiferromagnetic_2017} but questions remain about the exact processes of domain rearrangement and the associated energy scales and barriers \cite{meinert_electrical_2018, olejnik_antiferromagnetic_2017, jungfleisch_perspectives_2018}. 
In antiferromagnet-based spintronics, N\'{e}el vector switching and anisotropic magnetoresistance (AMR) can potentially be used to write and read data, respectively \cite{jourdan_epitaxial_2015, bodnar_magnetoresistance_2020, wadley_electrical_2016}.

%%%%%%%%%%%%%%%%%%%%%%%%%%%%%%%%%%%%%%%%%%%%%%%%%%%%%%%%%%%%%%%%%%%%%%%%%%%%%%%%%%%%%%%%%
% The structure of Mn$_2$Au
\begin{figure}
\centering
\includegraphics[width=0.9\columnwidth]{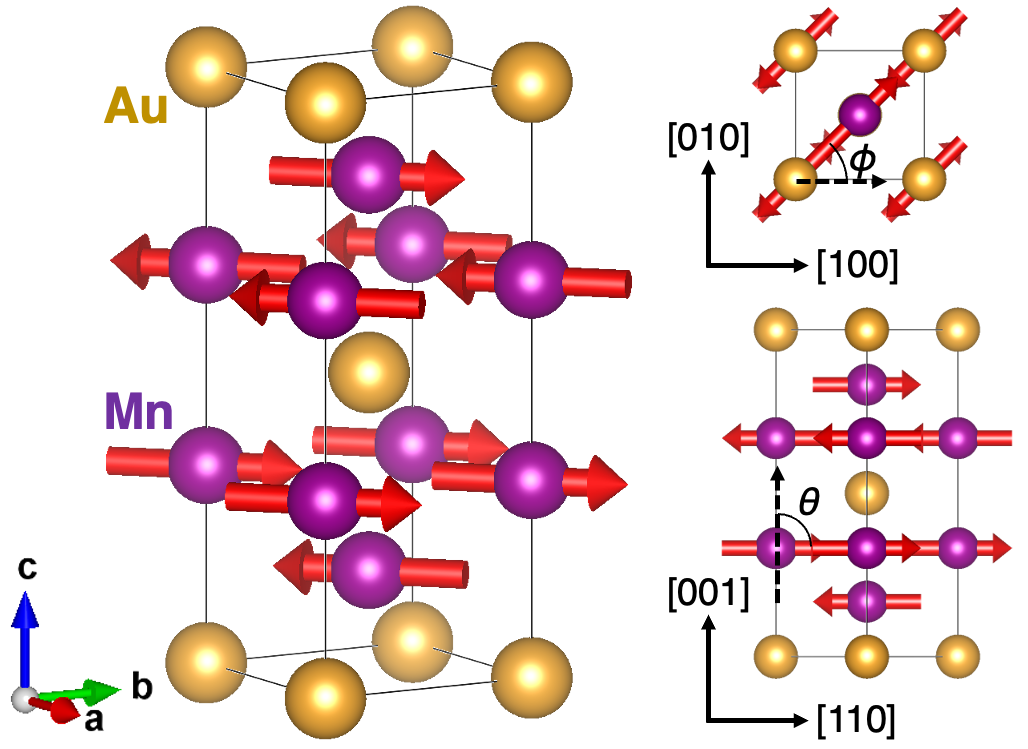}
\caption{\label{fig:structure}
Crystal structure of tetragonal Mn$_2$Au ($I4/mmm$, $a=3.33$~\AA, $c=8.54$~\AA) showing the antiferromagnetic  ordering with magnetic moments along $[110]$, the assumed easy axis from Sapozhnik, et al.\cite{sapozhnik_direct_2018}
The right side displays the $(001)$ plane and the $(1\bar{1}0)$ plane, while the angles $\theta$ and $\phi$ represent the angles utilized in our calculation of magnetocrystalline anisotropy.
}
\end{figure}
%%%%%%%%%%%%%%%%%%%%%%%%%%%%%%%%%%%%%%%%%%%%%%%%%%%%%%%%%%%%%%%%%%%%%%%%%%%%%%%%%%%%%%%%%

Mn$_2$Au is one of few room-temperature metallic antiferromagnets, along with CuMnAs and Cr$_2$Al, that have the crystal symmetry and collinear antiferromagnetic order that allows a staggered (Néel) accumulation of spins in the AF sublattices,\cite{bodnar_writing_2018, zhao_-plane_2021}
which in principle allows the electronic manipulation of the Néel vector using Néel spin-orbit torques. \cite{siddiqui_metallic_2020, wadley_electrical_2016}
{
\color{black} Mn$_2$Au has shown domain orientation by current-induced Néel vector switching,\cite{reimers2023current} though  the applied currents are large and the resistance can display a cumulative, sawtooth-shaped pattern rather than binary switching.\cite{bodnar_writing_2018}
}
The unit cell of Mn$_2$Au is shown in Fig.\ \ref{fig:structure} with arrows on Mn atoms to show this magnetic ordering.
Neutron diffraction and magnetization studies on powder Mn$_2$Au by Barthem \emph{et al.}\ confirm the $k=0$ antiferromagnetic structure of Mn$_2$Au with spins in the $ab$-plane.\cite{barthem_revealing_2013} 
{\color{black}
However, zero-field neutron diffraction 
}
of a tetragonal compound does not permit unique identification of the easy axis within the $ab$-plane. The powder study also showed an approximate susceptibility of $5 \times 10^{-4}$ averaged over all crystal directions. Neutron data showed a Mn moment of about 4~$\mu_B$, which is retained to about 85\,\% of its base value at 900~K \cite{barthem_revealing_2013}. 
Mn$_2$Au thus maintains its magnetic ordering until it decomposes upon heating into MnAu and an fcc solid solution phase, with a N\'{e}el temperature theorized up to 1300-1600 K \cite{shick_spin-orbit_2010, barthem_revealing_2013}. 
The presence of gold in Mn$_2$Au should also lead to a significant contribution from spin-orbit coupling.\cite{wu_anomalous_2016}
This high thermal stability is a hallmark of strong exchange interactions and anisotropy, and makes the material ideal for fundamental studies at room temperature or future data storage applications,\cite{bodnar_writing_2018}
but no bulk crystal growth has been reported to date.

Several studies of thin films and polycrystals have examined the magnetic and electrical properties of Mn$_2$Au. \cite{barthem_revealing_2013,bodnar_writing_2018,sapozhnik_direct_2018}
Thin film XMLD-PEEM studies by Sapozhnik \emph{et al.}\ found a magnetic easy axis of $\langle110\rangle$ and calculate an in-plane anisotropy term in the range 1\,--\,17 $\mu$eV/f.u., corresponding to lower and upper limits for the predicted spin flop fields H$_{SF}$ of 7 to 30 T.\cite{sapozhnik_direct_2018}
While the upper limit of 17 $\mu$eV/f.u.\ is evidenced by domain growth from  $\approx$1 $\mu$m along the equivalent [\,110]\, and [\,1$\bar{1}$0]\, easy axis at zero field to several microns perpendicular to the field at an applied field of 30 T, the lower limit was determined using domain wall widths near the resolution of the images.
Two studies by Bodnar \emph{et al.}\ \cite{bodnar_writing_2018, bodnar_magnetoresistance_2020} showed evidence of current-induced and magnetic-field induced Néel vector switching in Mn$_2$Au thin films separately. They also demonstrated AMR with a field along the hard $\langle100\rangle$ direction to be 6\,\%, which is much larger than that along the easy $\langle110\rangle$ crystal direction of $-0.15$\,\%.
The apparent switching behavior in magnetoresistance along $\langle110\rangle$ and $\langle100\rangle$ directions was discussed to be consistent with an extremely small/negligible in-plane anisotropy. 
This is consistent with an in-plane magnetocrystalline anisotropy energy (MAE) constant of 10 $\mu$eV/f.u.\ calculated by Shick \emph{et al.}\cite{shick_spin-orbit_2010}
{\color{black}
The effects of strain and defects are also apparent in a recent XMLD-PEEM study of patterend CuMnAs and Mn$_2$Au.\cite{reimers2024magnetic}
}

While most recent work has focused on Mn$_2$Au thin films, the fundamental properties of the material remain to be seen in bulk crystal form.
Notably, anisotropy may be strongly influenced by large stresses in epitaxial films and texturing from high applied currents.
In order to fully understand the material properties of Mn$_2$Au, we sought to synthesize and characterize bulk single crystals. 
According to the Mn-Au phase diagram, Mn$_2$Au decomposes into solid MnAu and Mn phases at 680$^{\circ}$\,C before melting \cite{massalski_au-mn_1985}. 
{\color{black}
This makes single crystal growth by directly cooling from the melt unfeasible. 
To grow a bulk single crystal, a flux method can be used to precipitate Mn and Au at lower temperatures.
}
Considering its binary phase diagrams with Au and Mn each, Bi was the most logical flux candidate \cite{cui_thermal_2014, okamoto_aubi_1983}. 
Here we describe the flux synthesis and physical characterization of the obtained crystals.  

%%%%%%%%%%%%%%%%%%%%%%%%%%%%%%%%%%%%%%%%%%%%%%%%%%%%%%%%%%%%%%%%%%
%Methods
\section{\label{sec:methods}Methods}

% The Quench Diagram
\begin{figure}
\centering
\includegraphics[width=0.99\columnwidth]{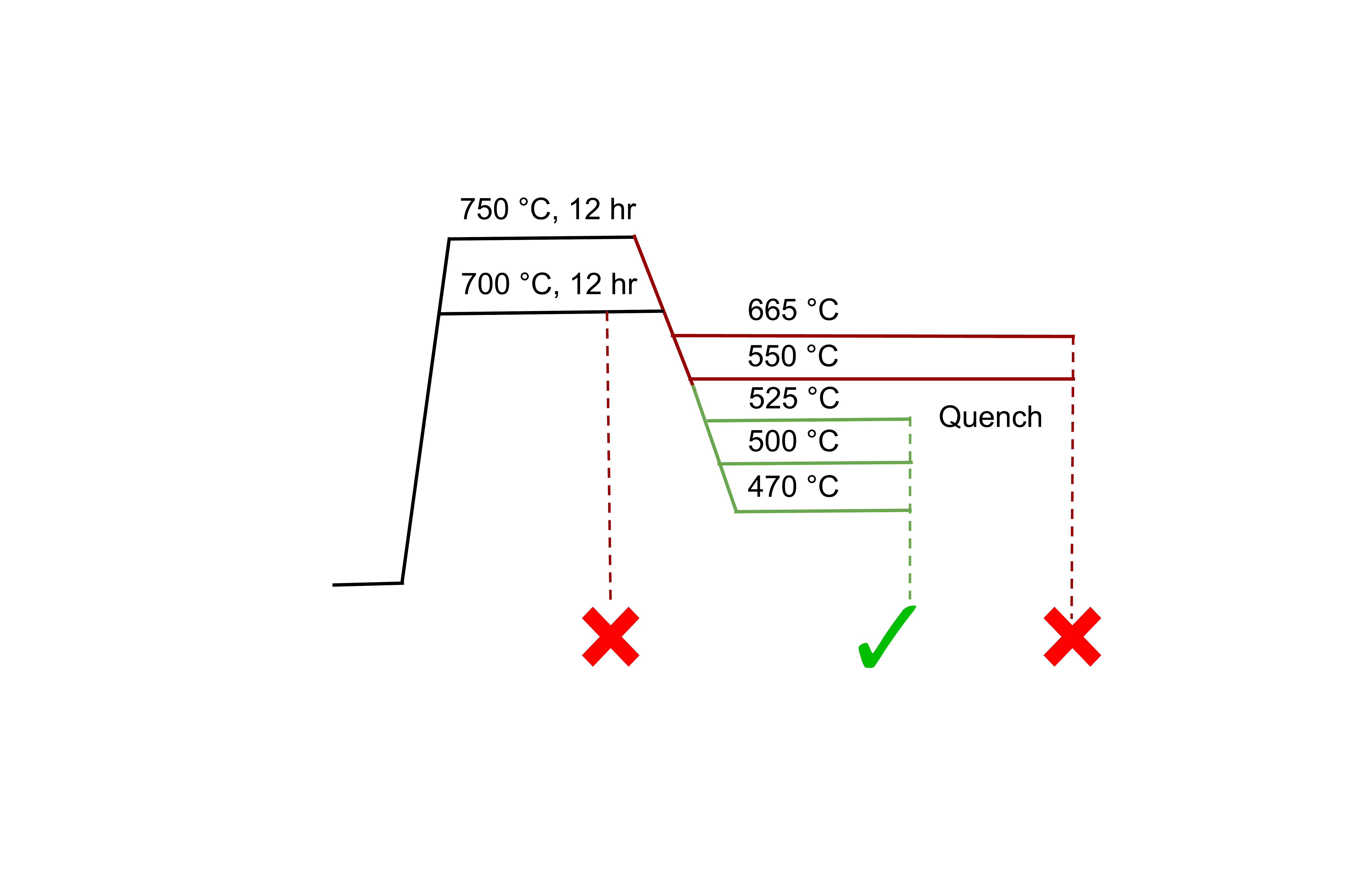}
\caption{\label{fig:quench}
Quenching experiments performed at a range of temperatures are summarized along with the respective phases formed. MnAu formation is represented by the red $\times$, Mn$_2$Au formation is represented by the green check.}
\end{figure}

\begin{figure}
\centering
\includegraphics[width=0.99\columnwidth]{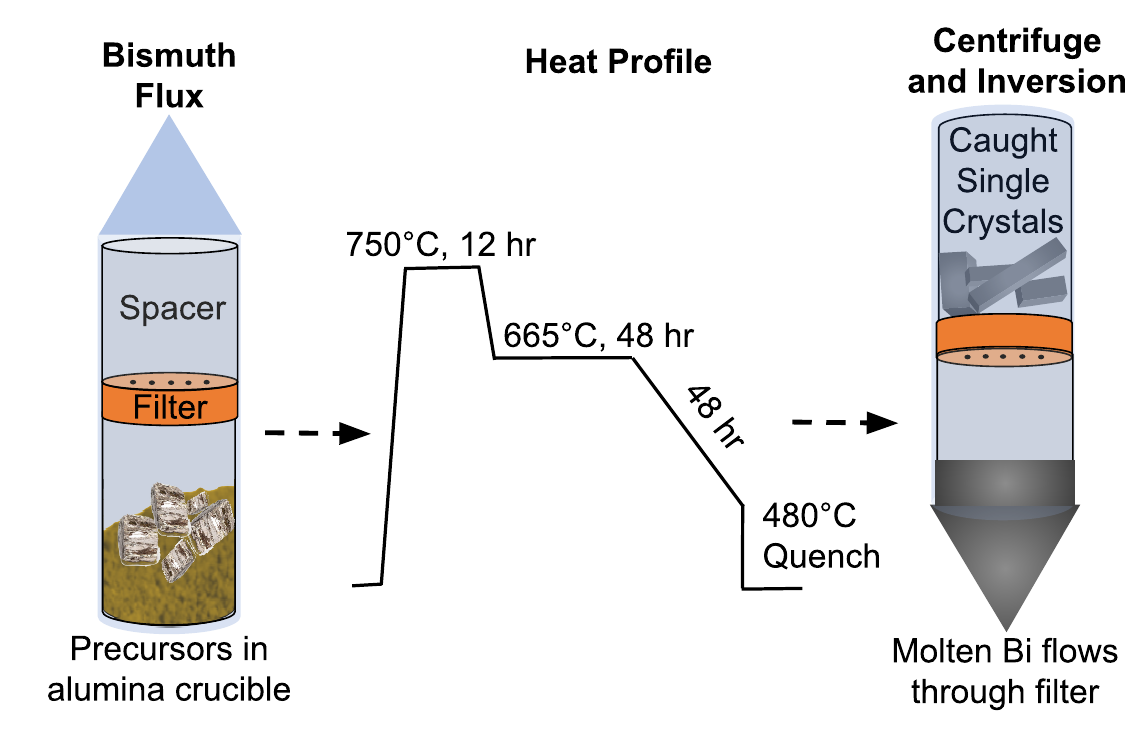}
\caption{\label{fig:flux}
Schematic of Mn$_2$Au flux crystal growth. Elemental precursors are loaded in an alumina crucible. The arrangement in a vacuum-sealed quartz ampoule allows the excess flux to be removed by decanting while crystals are caught by a fritted disk.}
\end{figure}

%\textbf{Crystal growth:}
Synthesis was performed using solid-state methods and metal flux techniques.
As-received Mn pieces with oxide-tarnished surfaces were cleaned by heating in vacuum at 1000$^{\circ}$\,C for 12 h, after which the Mn pieces are bright silver and the inner walls of the tube are dark.
Flux growth of Mn$_2$Au crystals was conducted using elemental precursors of Mn granules (99.98\,\%, 0.8\,--\,12 mm),  Au powder (99.99\,\%, 100 mesh), and   Bi chunks (99.99\,\%, $\le$15 mm) with an optimized nominal molar ratio of Mn:Au:Bi = 7:1:12.
The precursors were loaded into alumina crucibles, topped by alumina fritted disks\cite{canfield_use_2016} and a glass rod spacer in a fused silica ampoule. This setup, prepared in an argon-filled glovebox, was then evacuated and sealed at about 50 mTorr of pressure for reaction in an inert environment.
The precursors were heated to a maximum temperature of 750$^{\circ}$\,C for 12 h, quickly cooled and held at 650$^{\circ}$\,C for up to 24 hours below the dissociation temperature of Mn$_2$Au, slowly cooled to 470\,--\,480$^{\circ}$\,C for crystal growth, and from this temperature inverted and centrifuged to remove excess flux.
The synthesis is shown schematically in Fig.\ \ref{fig:flux}.
Variations of this recipe used to determine the precipitation temperature of the Mn$_2$Au phase and to optimize relative nominal amounts of precursors are summarized in Fig.\ \ref{fig:quench} and tabulated in the Supporting Information Table S1,\cite{supplement} along with a list of the largest crystal sizes obtained from each reaction.
Mn$_2$Au single crystals were manually isolated from remnant Bi flux by mechanically scraping or cleaving samples to isolate clean parts. Where remnant flux could not be easily removed mechanically, samples were acid etched with a 2:1 ratio of HNO$_3$ and HCl for less than 2 minutes. 

%%%%%%%%%%%%%%%%%%%%%%%%%%%%%%%%%%%%%%%%%%%%%%%%%%%%%%%%%%%%%%%%%%%%%%%%%%%%%%%%%%%%%%%%%
% The flux synthesis
%%%%%%%%%%%%%%%%%%%%%%%%%%%%%%%%%%%%%%%%%%%%%%%%%%%%%%%%%%%%%%%%%%%%%%%%%%%%%%%%%%%%%%%%%

%\textbf{Characterization:}
Powder X-ray diffraction was conducted in transmission geometry with a Bruker D8 diffractometer with Mo-K$\alpha$ radiation.
Reflection X-ray diffraction was performed with a Bruker D8 diffractometer with Cu-K$\alpha$ radiation on single-crystalline samples to confirm crystal orientation. Rietveld analysis was performed using GSAS-II \cite{toby_gsas-ii_2013}.

Microstructure and composition were assessed using a ThermoFisher Axia ChemiSEM scanning electron microscope with electron dispersive X-ray spectroscopy (SEM-EDS) using an acceleration voltage of 20 kV.
A Shimadzu Differential Thermal Analyzer (DTA-50) was used to examine the thermal stability of the flux-synthesized Mn$_2$Au single crystals between room temperature and 750$^{\circ}$\,C, with samples in sealed quartz ampoules.
Electron backscatter diffraction (EBSD) mapping was conducted in a Thermo Scios2 Dual-Beam SEM/FIB instrument at an acceleration voltage of 30 kV. 
Magnetic susceptibility measurements were performed using a Quantum Design Magnetic Property Measurement System (MPMS).
Four-point resistivity measurements were made using a Quantum Design Physical Property Measurement System (PPMS).

Single-crystal neutron diffraction was collected on a $1\times0.1\times0.1$~mm$^3$ Mn$_2$Au crystal on the HB-3A four-circle diffractometer at the High Flux Isotope Reactor (HFIR) at Oak Ridge National Laboratory. A neutron wavelength of 1.542 \AA \ was used and measurements were conducted at 295 K. Magnetic symmetry analysis was done using the MAXMAGN tool in the Bilbao Crystallographic Server \cite{perez-mato_symmetry-based_2015}. Single crystal neutron diffraction refinements of the nuclear and magnetic structure were done using the FULLPROF suite \cite{rodriguez-carvajal_recent_1993}.

First-principles density functional theory (DFT) simulations were carried out using the Vienna \emph{Ab-Initio} Simulation Package (\texttt{VASP}), employing the projector-augmented-wave (PAW) formalism \cite{Kresse:1996, Kresse:1999} to describe the electron-ion interaction.
Kohn-Sham states were expanded into plane waves with a cutoff energy of 600 eV.
The exchange-correlation term in the Kohn-Sham equation was described using the Perdew, Burke, and Ernzerhof (PBE) formulation of the generalized-gradient approximation\cite{Perdew:1997}.
We also used the DFT+$U$ approach\cite{Dudarev:1998} and selected an effective on-site Coulomb interaction parameter $U_{\mathrm{eff}}$=1.4\,eV, resulting in good agreement between lattice parameters from our experiment and those from another X-ray diffraction experiment \cite{Wells:1970}, see details in Fig.\ S10 in the supplemental information.
The measured magnetic moments from powder samples \cite{barthem_revealing_2013} further substantiate the suitability of our choice of $U_{\mathrm{eff}}$ value.
However, the magnetic moments derived from this work are about 0.5 $\mu_B$ larger, see Fig.\ S10 in the supplemental information.
For a reliable representation of the in-plane and out-of-plane magnetocrystalline anisotropy energy, a Monkhorst-Pack $\mathbf{k}$-point mesh \cite{Monkhorst:1976} of $24\times24\times9$ was meticulously chosen, as depicted in Fig.\ S8 of the supplemental information.
We ensure $\mathbf{k}$-point convergence of the anisotropic energy difference within 540 kJ/m$^{3}$ for $K_{1}$ and 5 kJ/m$^{3}$ for $K_{22}$.
%All self-consistent calculations considered the influence of spin-orbit coupling within the non-collinear magnetism framework by applying a non self-consistent field (SCF) calculation for the spin-orbit coupling effect after the spin-polarized SCF calculation \cite{Steiner:2016}.
The influence of spin-orbit coupling is included within the non-collinear magnetism framework by applying a non self-consistent field (SCF) calculation for the spin-orbit coupling effect after a spin-polarized SCF calculation \cite{Steiner:2016}.
Atomic geometries are relaxed using the SCF spin-polarized approach only, i.e.\ without spin-orbit coupling.
In all cases, the magnetic structure was included in the description.
In the calculation of magnetic susceptibility, we enforced the angular constraint on the magnetic moments using a penalty term in the total energy, as implemented in \texttt{VASP}.

%%%%%%%%%%%%%%%%%%%%%%%%%%%%%%%%%%%%%%%%%%%%%%%%%%%%%%%%%%%%%%%%%%%%%%%%%%%%%%%%%%%%%%%%%%%%%%%%%%%%%%%%%%%%%%%%%%%%%%%%%%%%%%%%%%%%%%%%%%%%%%%%%%%%%%%%%%%%%%%%%%%%%%%%%%%%%%%%%%%%%%%%%%%%%%%%%%%%%%%%%%%%%%%%%%%%%%%%%%%%%%%%%%%%%%%%%%%%%%%%%%%%%%%%%%%%%%%%%%%%%%%%%%%%%%%%%%%%%%%%%%%%%%%%%%%%%%%%%%%%%%%%%%%%%%%%%%%%%%%%%%%%%%%%%%%%%%%%%%%%%%%%%%%%%%%%%%%%%%%%%%%%%%%%%%%%%%%%%%%%%%%%%%%%%%%%%%%%%%%%%%%%%%%%%%%%%%%%%%%%%%%%%%%%%%%%%%%%%%%%%%%%%%%%%%%%%%%%%%%%%%%%%
%%%RESULTS and DISCUSSION%%%
\section{Results and Discussion}

Bulk single crystals of Mn$_2$Au, with dimensions on the mm-scale, were grown using Bi flux.
The decomposition temperature of Mn$_2$Au in the solid state is reported to be 680$^{\circ}$\,C \cite{freund_thin_2004}.
In bismuth flux, our quenching experiments, summarized in Fig.\ \ref{fig:quench}, reveal that Mn$_2$Au precipitated from flux below 550$^{\circ}$\,C, while MnAu formed between 550 and 700$^{\circ}$\,C. 

Differential thermal analysis (DTA) of isolated Mn$_2$Au crystals was used to assess the dissociation temperature compared to the existing Mn--Au phase diagram\cite{boucher_coordination_1989}.
The DTA signal for pure Mn$_2$Au crystals, Fig.\ S3 in the Supporting Information, \cite{supplement} shows phase changes with onset at $\sim$680$^{\circ}$\,C on heating and $\sim$630$^{\circ}$\,C on cooling.
This sets the upper temperature bound for applications using Mn$_2$Au, as heating a spintronic device past 630$^{\circ}$\,C could destroy the Mn/Au ordering.

%%%%%%%%%%%%%%%%%%%%%%%%%%%%%%%%%%%%%%%%%%%%%%%%%%%%%%%%%%%%%%%%%%%%%%%%%%%%%%%%%%%%%%%%%
%%%%%%%%%%%%%%%%%%%%%%%%%%%%%%%%%%%%%%%%%%%%%%%%%%%%%%%%%%%%%%%%%%%%%%%%%%%%%%%%%%%%%%%%%

%%%%%%%%%%%%%%%%%%%%%%%%%%%%%%%%%%%%%%%%%%%%%%%%%%%%%%%%%%%%%%%%%%%%%%%%%%%%%%%%%%%%%%%%%%
% Optical pictures
\begin{figure}
\centering
\includegraphics[width=0.9\columnwidth]{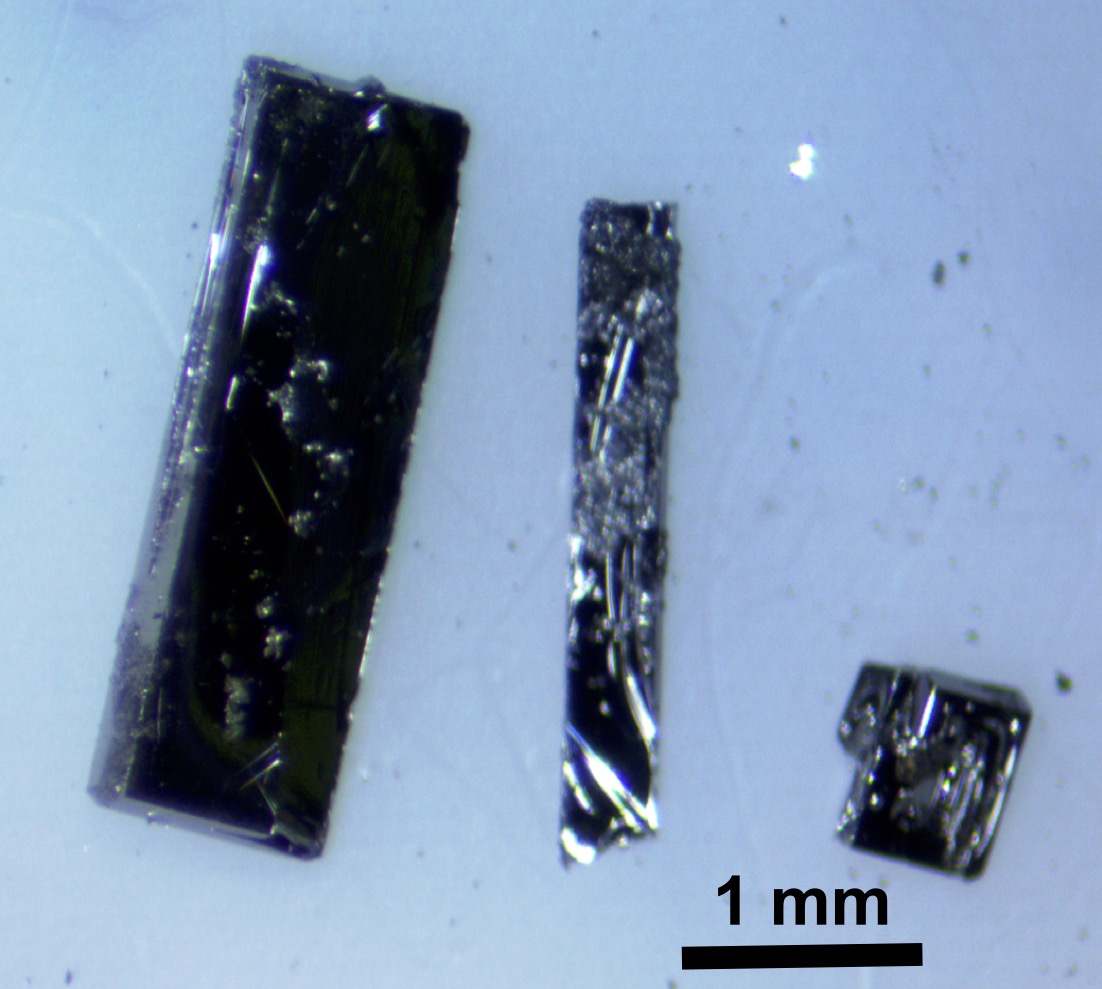}
\caption{\label{fig:optical}
Optical micrograph of representative Mn$_2$Au bulk crystals with plate, needle, and square morphologies.}
\end{figure}
%%%%%%%%%%%%%%%%%%%%%%%%%%%%%%%%%%%%%%%%%%%%%%%%%%%%%%%%%%%%%%%%%%%%%%%%%%%%%%%%%%%%%%%%%% 

In order to find optimal heating and composition protocols, many samples were prepared under varying conditions such as the nominal amounts of Mn:Bi and the heating profile. A variety of crystal shapes were obtained, as shown in Fig.\ \ref{fig:optical}, including cubes, thick needles, and flat plates on the mm-scale. There may be some significance of specific heating steps that were attempted:
As an example, a hold at 650$^{\circ}$\,C seems to favor a smaller quantity of larger single crystals, and larger Mn:Bi ratios favor plate-like morphology over thin needles.
However, a mechanistic understanding of these steps remains to be fully developed due to the complexity of the process.
Variations of flux syntheses are tabulated in the Supporting Information\cite{supplement}.
The largest crystals at optimized conditions were used for further characterization.

%%%%%%%%%%%%%%%%%%%%%%%%%%%%%%%%%%%%%%%%%%%%%%%%%%%%%%%%%%%%%%%%%%%%%%%%%%%%%%%%%%%%%%%%%%
%typical xrd
\begin{figure}
\centering
\includegraphics[width=0.95\columnwidth]{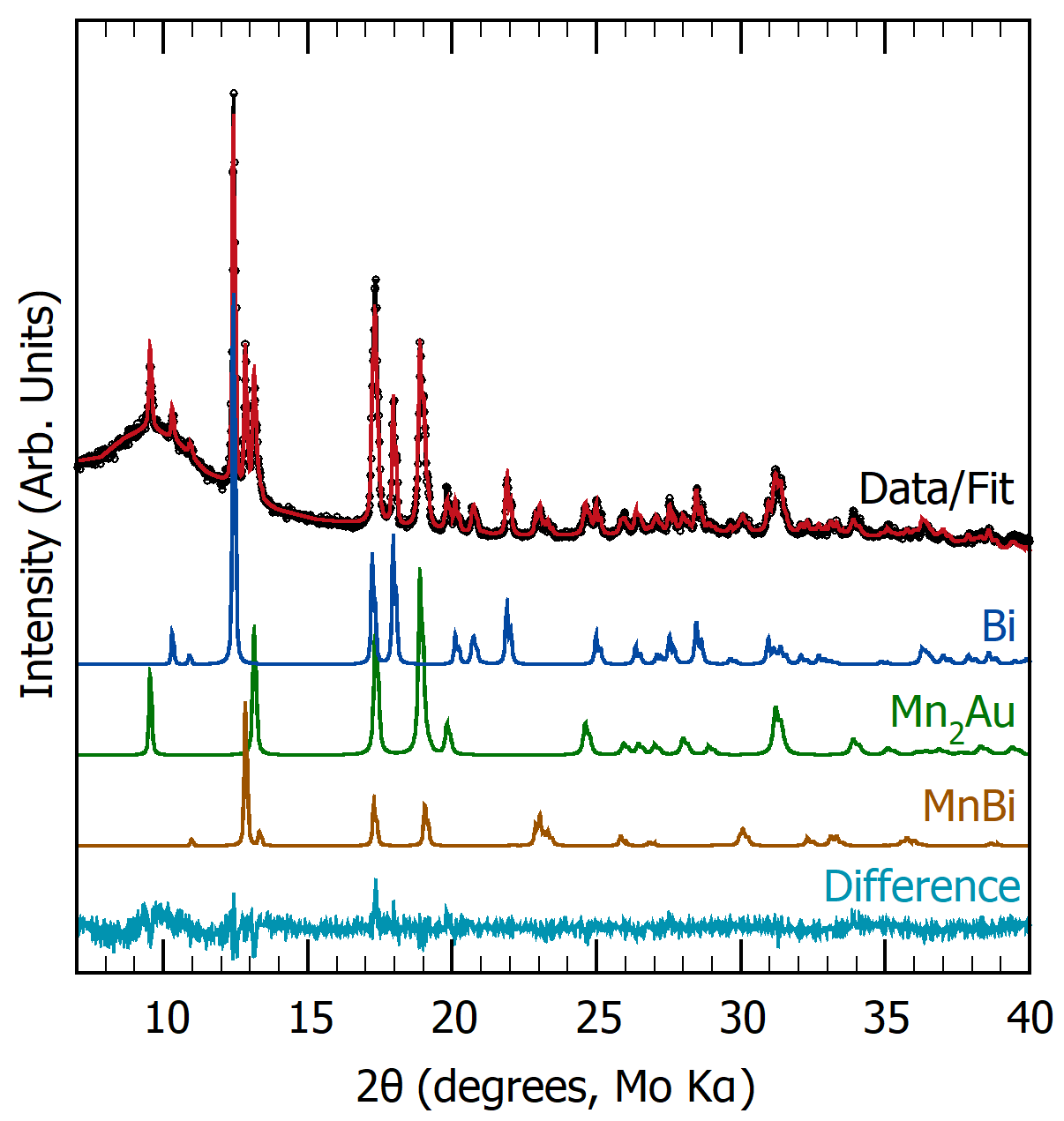}
\caption{\label{fig:pxrd}
Powder XRD and Rietveld-refined contribution of a typical Bi-flux growth that yields the desired Mn$_2$Au phase along with Bi and MnBi impurities.
Bi and MnBi are removed prior to magnetic characterization, as described below. }
\end{figure}
%%%%%%%%%%%%%%%%%%%%%%%%%%%%%%%%%%%%%%%%%%%%%%%%%%%%%%%%%%%%%%%%%%%%%%%%%%%%%%%%%%%%%%%%%

Rietveld refinement of powder XRD for Bi-flux products caught after centrifuging is presented in Fig.\ \ref{fig:pxrd}.
The refinement shows 58 mol $\%$ Mn$_2$Au, 20 mol $\%$ Bi, and 22 mol $\%$ MnBi.
This distribution indicates that the Mn$_2$Au single crystals have residual MnBi and Bi on their surfaces.
The Mn$_2$Au single crystals used for further characterizations were picked out and cleaned mechanically.
However, a small amount of Bi and MnBi typically remains on the surface, which is subsequently removed (to the best of our ability) by etching as described in Sec.\ \ref{sec:methods}.
This phase distribution in the reaction products is visually apparent in SEM-EDS images shown in Fig.\ S1 \cite{supplement}.

%%%%%%%%%%%%%%%%%%%%%%%%%%%%%%%%%%%%%%%%%%%%%%%%%%%%%%%%%%%%%%%%%%%%%%%%%%%%%%%%%%%%%%%%%

%%%%%%%%%%%%%%%%%%%%%%%%%%%%%%%%%%%%%%%%%%%%%%%%%%%%%%%%%%%%%%%%%%%%%%%%%%%%%%%%%%%%%%%%%
% The  Reflection plot
\begin{figure}
\centering
\includegraphics[width=1\columnwidth]{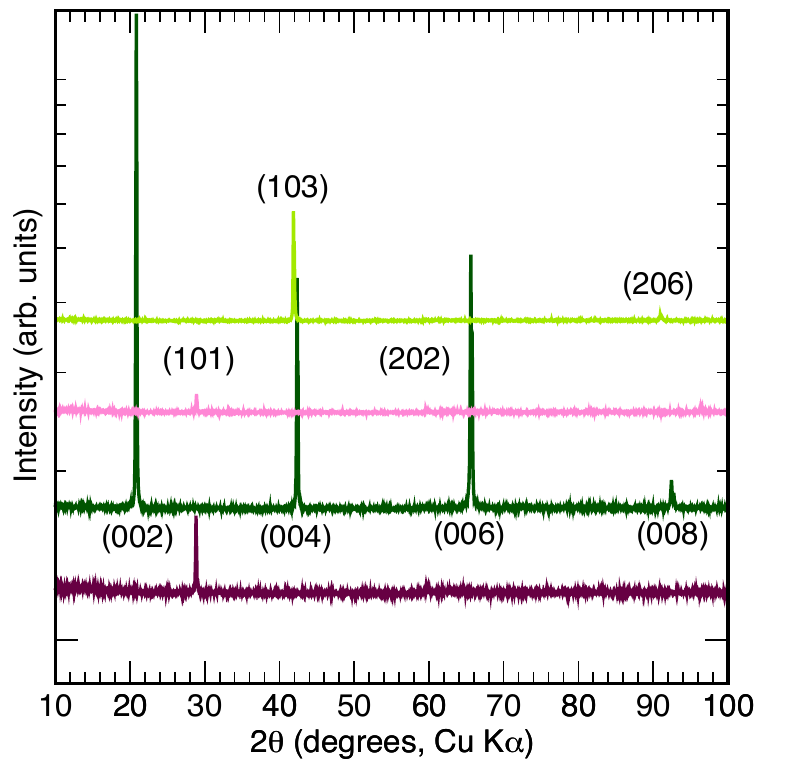}
\caption{\label{fig:refXRD}
Reflection XRD of representative Mn$_2$Au crystals with different morphologies.
Light and dark pink curves are data from needle-like samples where the direction normal to long needle axis is shown to have (101) character. 
Light and dark green curves collected from the large exposed faces of plate-like samples. 
The light green sample shows an oblique (103) orientation, while the dark green curve shows that the $c$-axis points out of the face. This (001) face orientation is the most commonly observed orientation of flat crystals. 
}
\end{figure}
%%%%%%%%%%%%%%%%%%%%%%%%%%%%%%%%%%%%%%%%%%%%%%%%%%%%%%%%%%%%%%%%%%%%%%%%%%%%%%%%%%%%%%%%%
Reflection XRD (see Fig.\ \ref{fig:refXRD}) is used as a preliminary tool to check the orientation of Mn$_2$Au single crystals.
The diffraction data in green is obtained for plate-like crystals, showing that the (001) plane is the largest exposed plane.
The data in pink were obtained for crystals with horizontally lying needle-like crystal formation; the reflections correspond to $\langle h0l \rangle$.
The large plates have the $c$-axis pointing perpendicular to the widest exposed plane while the narrow needles have the $c$-axis lying in the plane perpendicular to the longest sample dimension.

%%%%%%%%%%%%%%%%%%%%%%%%%%%%%%%%%%%%%%%%%%%%%%%%%%%%%%%%%%%%%%%%%%%%%%%%%%%%%%%%%%%%%%%%%
% The EBSD map
\begin{figure}
\centering
\includegraphics[width=0.99\columnwidth]{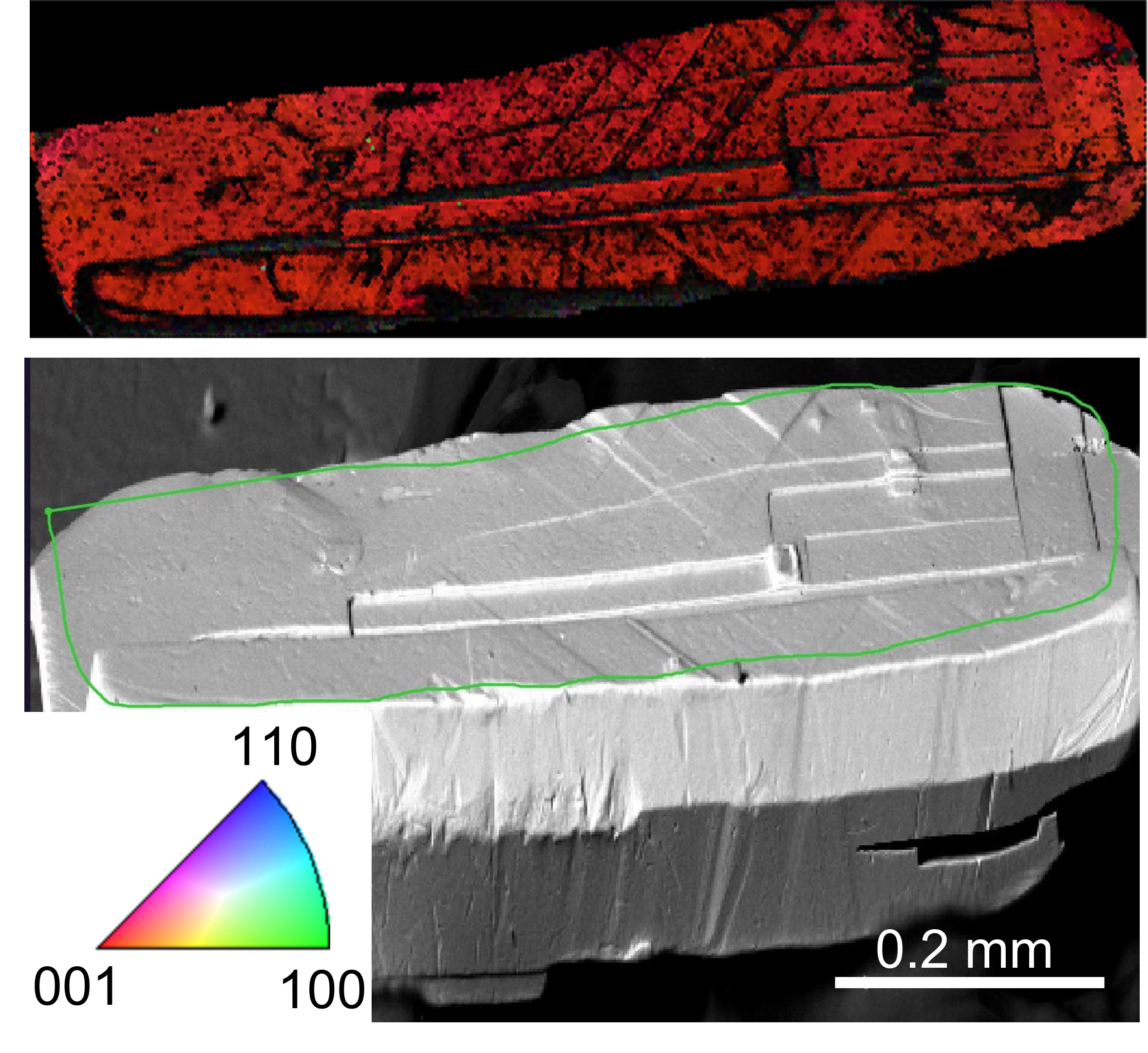}
\caption{\label{fig:EBSD}
Electron backscatter diffraction (EBSD) map (top) and corresponding electron micrograph (bottom) confirm that the Mn$_2$Au $c$-axis is normal to the large flat crystal face. An inset shows the inverse pole figure key where red corresponds to the (001) lattice plane.}
\end{figure}
%%%%%%%%%%%%%%%%%%%%%%%%%%%%%%%%%%%%%%%%%%%%%%%%%%%%%%%%%%%%%%%%%%%%%%%%%%%%%%%%%%%%%%%%%

Electron backscatter diffraction (EBSD) maps as in Fig.\ \ref{fig:EBSD} were used to confirm the orientation of mm-scale plate-like Mn$_2$Au crystals.
The map shows that the $c$-axis of the crystal is perpendicular to the largest exposed sample plane. The inverse pole figure shows that the entire face is red, corresponding to 001 orientation. 
Since EBSD is quite surface-sensitive, crystal facets and steps with different orientations appear dark as their Kikuchi patterns are not well-recorded. The EBSD data also offers 3-dimensional information of orientation within the surface plane. The vector difference between two points along the longest dimension (horizontal in Fig.\ \ref{fig:EBSD}) corresponds to the 100 crystal direction. Accordingly, the diagonal dark lines with 110 orientation are growth terraces. 
Anisotropic magnetic and transport measurements on this and other Mn$_2$Au crystals were performed after confirming the orientation of each with XRD and EBSD.

\begin{figure}
\centering
\includegraphics[width=0.85\columnwidth]{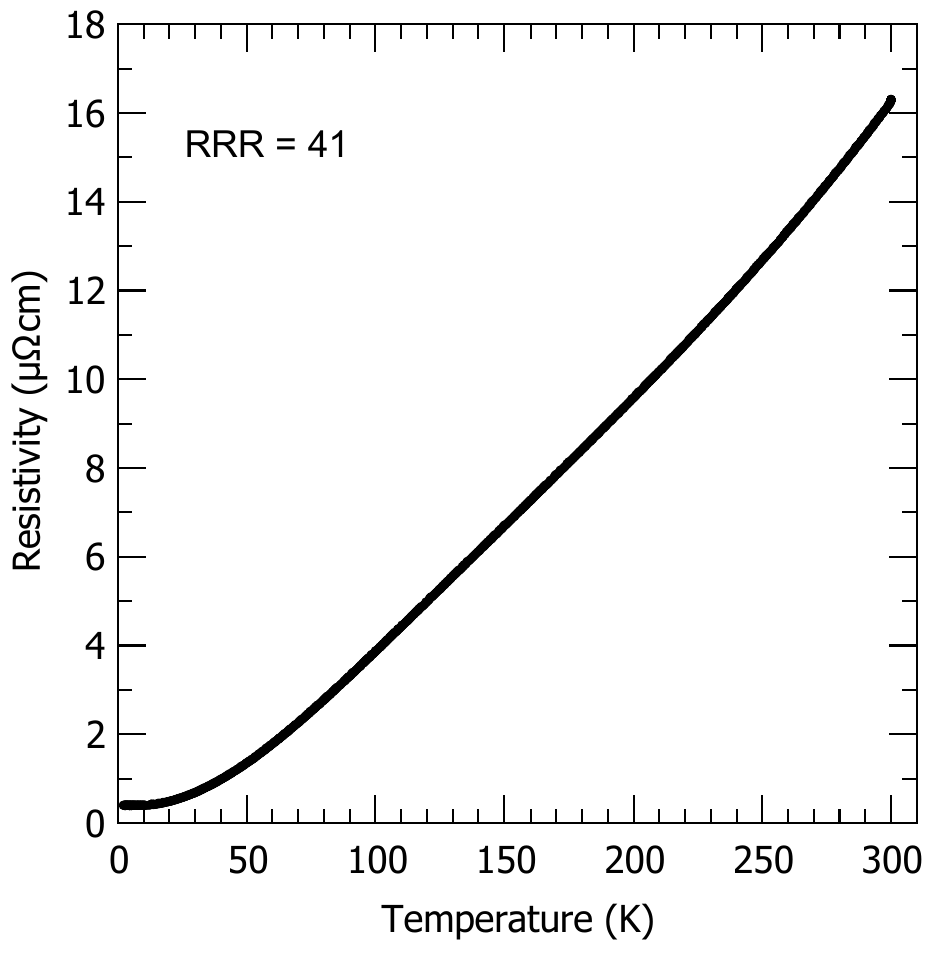}
\caption{\label{fig:resistivity}
Four-point resistivity of a bulk single crystal of Mn$_2$Au along $\langle 100 \rangle$ shows metallic behavior with a residual resistivity ratio (RRR) of 41.}
\end{figure}

Four-point resistivity measurements with current applied along the length of a needle-like crystal along $\langle 100 \rangle$ confirm the metallic behavior of Mn$_2$Au with a room temperature resistivity of $\sim 16$ $\mu\Omega$cm (see Fig.\ \ref{fig:resistivity}) and a low temperature (2 K) resistivity of $\sim 0.4$ $\mu\Omega$cm. This compares to resistivity of $\sim 21$ and $\sim 7$ $\mu\Omega$cm reported for an epitaxial film at 300 K and 2 K respectively \cite{jourdan_epitaxial_2015}. The residual resistivity ratio (RRR) of 41 in our bulk sample is an order of magnitude larger than that reported for epitaxial films \cite{jourdan_epitaxial_2015} because of the small low-temperature resistivity in the bulk sample. 

Similar four-point resistivity measurements along other crystal directions are limited by the sample's aspect ratio. We attempted to conduct measurements of micron-scale lamellae by focused ion beam liftout, but a high concentration of implanted Ga led to a drastic decrease in the RRR.

%%%%%%%%%%%%%%%%%%%%%%%%%%%%%%%%%%%%%%%%%%%%%%%%%%%%%%%%%%%%%%%%%%%%%%%%%%%%%%%%%%%%%%%%%%
% The resistivity plot
%%%%%%%%%%%%%%%%%%%%%%%%%%%%%%%%%%%%%%%%%%%%%%%%%%%%%%%%%%%%%%%%%%%%%%%%%%%%%%%%%%%%%%%%%

%%%%%%%%%%%%%%%%%%%%%%%%%%%%%%%%%%%%%%%%%%%%%%%%%%%%%%%%%%%%%%%%%%%%%%%%%%%%%%%%%%%%%%%%%
% Single crystal neutron refinement with Neel vector along 100 vs 110 are comparable (can't independently solve anisotropy question)
\begin{figure}
\centering
\includegraphics[width=0.9\columnwidth]{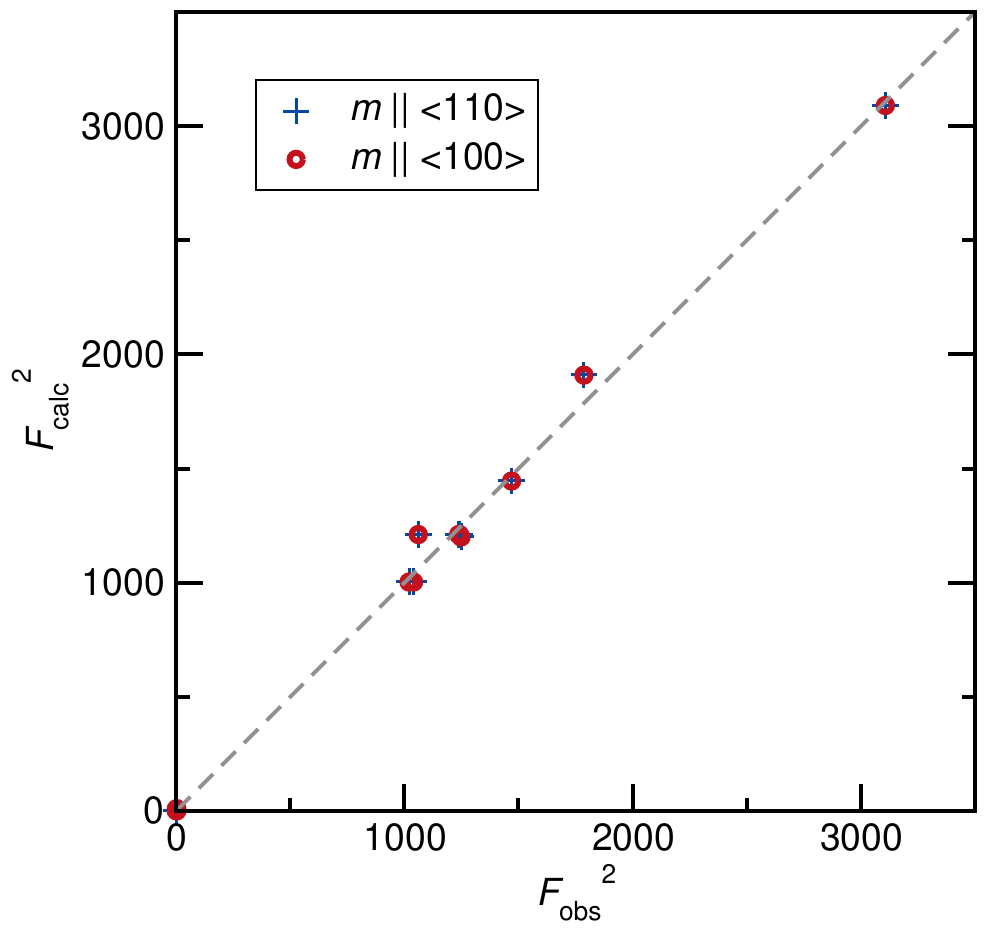}
\caption{\label{fig:Neutron}
Calculated versus observed structure factors for a phase of Mn$_2$Au that includes both magnetic and nuclear contributions are obtained from single crystal neutron diffraction.
Two models with N\'{e}el vectors along the $\langle 110 \rangle$ and $\langle 100 \rangle$ family of directions confirm in-plane moments but give equivalent fits that cannot reveal the in-plane moment direction.
That direction is determined by magnetic susceptibility measurements. 
}
\end{figure}
%%%%%%%%%%%%%%%%%%%%%%%%%%%%%%%%%%%%%%%%%%%%%%%%%%%%%%%%%%%%%%%%%%%%%%%%%%%%%%%%%%%%%%%%%

Single crystal neutron diffraction data refinements (see Fig.\ \ref{fig:Neutron}) show excellent agreement between calculated and observed structure factors for antiferromagnetic tetragonal Mn$_2$Au with a propagation vector $k = [000]$. Refinements were carried out using two different magnetic space groups corresponding to the two possible Néel vectors within the $ab$-plane for the commensurate AF phase in the nuclear $I4/mmm$ space group. The $Fm'mm$ magnetic space group (\#69.523) was used to refine the AF phase with a Néel vector along $\langle 110 \rangle$.
The $Im'mm$ magnetic space group (\#71.535) was used to refine that with the Néel vector along $\langle 100 \rangle$.
Both refinements converge to a Mn moment of 4.56(2) $\mu_B$, which is slightly larger than the DFT-obtained value of 3.98 $\mu_B$ and the one measured on powder samples of 4 $\mu_B$ \cite{barthem_revealing_2013}. 
Visually, it is apparent from Fig.\ \ref{fig:Neutron} that both models describe the experimental data equally well.
Both fits give $R_F$ = 0.019 and $\chi^2=0.26$, due to the symmetry equivalence of the two in-plane models for single-crystal neutron scattering.
Since the easy axis cannot be distinguished between $\langle 110 \rangle$ and $\langle 100 \rangle$,
a complementary confirmation of in-plane anisotropy is presented using the directional magnetic property measurements that follow.

Magnetization isotherms and temperature sweeps were used to characterize the magnetic response of samples over a large span of temperatures between 10 K and 700 K, but careful experimentation is required to separate the intrinsic response of Mn$_2$Au from dilute MnBi impurities. 
Magnetization isotherms at 300 and 700 K, as well as intermediate magnetization versus temperature scans, discussed in detail in the Supplemental Information \cite{supplement}, revealed that amounts as low as 2 mol \% of the MnBi-low temperature phase (LTP) dominate the magnetic response.
LTP-MnBi orders ferromagnetically at 630 K \cite{mcguire_symmetry-lowering_2014}, concurrent with its phase change to the high temperature phase (HTP). Room temperature scans are easily overwhelmed by the ferromagnetic contribution from the MnBi impurity exhibiting a large magnetization $M$, a non-linear saturation of $M$, and hysteresis. Above the ordering temperature of LTP-MnBi, however, $M$ is linear with applied field $H$ with a susceptibility on the order of 10$^{-4}$ as expected for Mn$_2$Au.\cite{barthem_revealing_2013}
The disappearance of ferromagnetic behavior above the Curie temperature of the MnBi impurity confirms that the FM observed at room temperature is not an intrinsic property of the Mn$_2$Au sample itself, but rather of the MnBi impurity. 

To obtain anisotropic susceptibility of the single crystals, this large FM contribution had to be reduced further. Reducing the ferromagnetic signal at room temperature, without having to break clean sections of samples, was explored via two ways. In the first approach inspired by high temperature magnetization measurements, a sample was heated to 700 K in vacuum and quenched quickly to avoid reformation of the LTP-MnBi phase. In the second approach, the Mn$_2$Au crystals were etched with a 1:2 HCl and HNO$_3$ solution for 2 minutes to remove the Bi flux. Whereas both helped reduce the FM contribution, they brought the risk of oxidizing the sample or leaching the Mn from the sample. Further details on these methods are discussed in the Supporting Information \cite{supplement} and references \citenum{andresen_magnetic_1967,taufour_structural_2015,hazek_dissolution_2016,lenher_solubility_2002,narayani_mechanism_2019} within.

%%%%%%%%%%%%%%%%%%%%%%%%%%%%%%%%%%%%%%%%%%%%%%%%%%%%%%%%%%%%%%%%%%%%%%%%%%%%%%%%%%%%%%%%%
% in and out of plane anisotropy at 10 K and 300 K
\begin{figure}
\centering
\includegraphics[width=0.99\columnwidth]{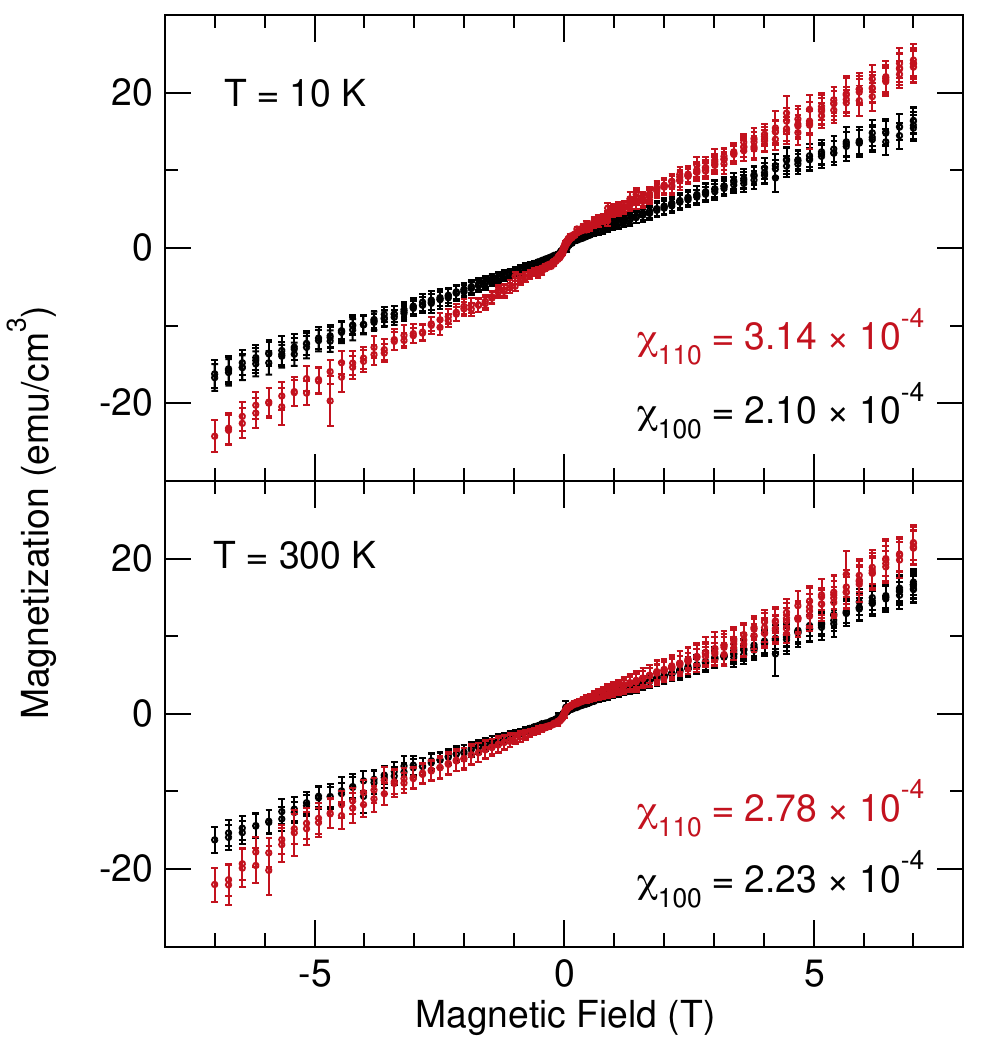}
\caption{\label{fig:Aniso}
Magnetization isotherms at 10 and 300 K with field applied along the $\langle 110\rangle$ and $\langle 100\rangle$ crystal directions show evidence of in-plane anisotropy. 
Susceptibilities extracted from the data with $H > 1$~T are listed.
The ratios of $\chi_{110}/\chi_{100}$  are 1.5 and 1.2, similar to $\sqrt{2}$.
}
\end{figure}
%%%%%%%%%%%%%%%%%%%%%%%%%%%%%%%%%%%%%%%%%%%%%%%%%%%%%%%%%%%%%%%%%%%%%%%%%%%%%%%%%%%%%%%%%

The cleanest Mn$_2$Au samples were obtained by manually cutting the flux-free sections of as-synthesized single crystals. Magnetic isotherms measured on this sample at 10 and 300 K with the field along $\langle 110 \rangle$ and $\langle 100 \rangle$ are shown in Fig.\ \ref{fig:Aniso}. A mostly linear behavior in $M$ with field $H$ is observed for all measurements with the exception of a steeper region between $\pm 0.1$ T. While the behavior near $H=0$~T may be due to trace amounts of MnBi, the predominantly linear behavior at higher fields represents the intrinsic Mn$_2$Au susceptibility. The dimensionless susceptibility $\chi$ is calculated using the slope of this linear section of the data for applied fields $H > 0.1$~T. The susceptibility in each case is on the order of 10$^{-4}$, matching the calculated value of 3.98$\times$10$^{-4}$ for  Mn$_2$Au in the literature \cite{kang_polar_2022}. The in-plane diagonal susceptibility $\chi_{110}$ is larger than $\chi_{100}$ by a factor of 1.5 and 1.25 at 10~K and 300~K, respectively.
Different samples, acid-cleaned by different methods, consistently showed that $\chi_{110} > \chi_{100}$, as discussed in the Supporting Information \cite{supplement}.
No subtle features that would indicate a polarization of in-plane domains or a spin-flop field are apparent in the linear regions of susceptibility for Mn$_2$Au, unlike those observed in the more susceptible in-plane antiferromagnet Fe$_2$As.\cite{yang_magnetocrystalline_2020}
As we discuss below, the in-plane spin flop field is predicted to be greater than 30~T.

% MPMS QD limits are on the order of 10$^{-7}$ emu

The magnetic susceptibility measurements in Fig.\ \ref{fig:Aniso} complement our neutron diffraction data by probing the in-plane N\'{e}el direction.
Mn$_2$Au is antiferromagnetic at both temperatures, 300~K and 10~K, and the susceptibility measurements show that the in-plane susceptibilities are approximately equal at room temperature, with a small anisotropy at 10~K.
These susceptibilities are very small, on the same order as the isostructural room-temperature AF Cr$_2$Al,\cite{zhao_-plane_2021} and about 100 times smaller than than Fe$_2$As.\cite{yang_magnetocrystalline_2020}
The susceptibility along the $c$-axis $\chi_{001}$ is about $4.9 \times 10^{-4}$, on the same order of magnitude as $\chi_{100}$ and $\chi_{110}$ (shown in Supplemental Information Figure S7).\cite{supplement}
For a tetragonal crystal, the low-field (below spin flop) in-plane susceptibility should be isotropic--there are only $\chi_{11}$, $\chi_{33}$, and $\chi_{13}$ terms in the susceptibility (as discussed in the Supplemental Information\cite{supplement} and in the texts by Nye\cite{nye_physical_1990} and Newhnam),\cite{newnham_properties_2005} so the measured difference we see between $\chi_{100}$ and $\chi_{110}$ is likely due to shape anisotropy of the Mn$_2$Au or texturing of minuscule amounts of ferromagnetic MnBi impurity (discussed in detail in the Supplemental Information).\cite{supplement}
The $\langle 110 \rangle$ direction was proposed to be the easy axis by thin film studies employing XMLD-PEEM for direct domain imaging \cite{sapozhnik_direct_2018}.
The thin film study shows easy-axis domain growth when fields above 30~T are applied perpendicular to $\langle 110 \rangle$.

%%%%%%%%%%%%%%%%%%%%%%%%%%%%%%%%%%%%%%%%%%%%%%%%%%%%%%%%%%%%%%%%%%%%%%%%%%%%%%%%%%%%%%%%%
% Calculated magnetocrystalline anisotropy results from DFT
\begin{figure}
\centering
\includegraphics[width=0.99\columnwidth]{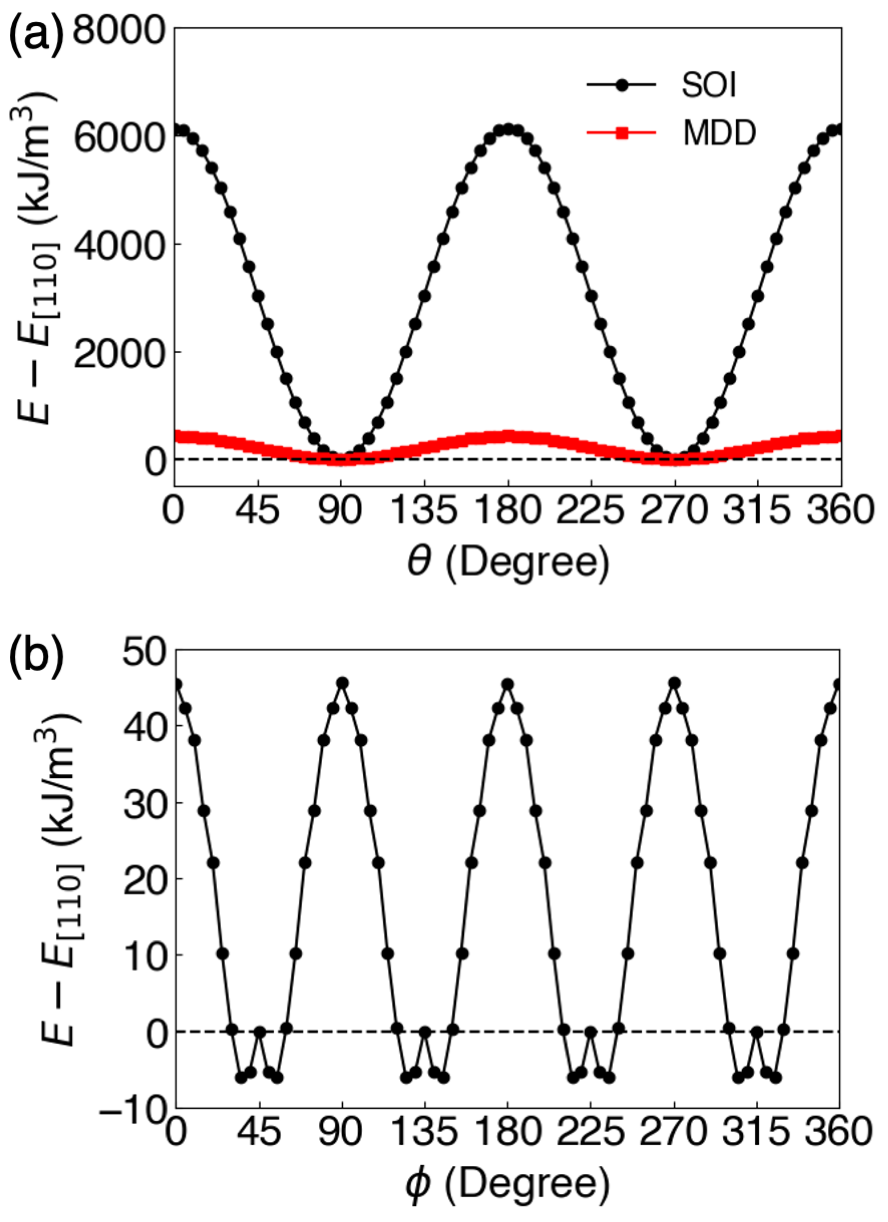}
\caption{\label{fig:Aniso-DFT}
Magnetocrystalline anisotropy of Mn$_{2}$Au from DFT calculations shows that (a) in the (110) plane, minimum energy ($E=0$) occurs when the spins point along $\langle 1\bar{1}0 \rangle$, which confirms that (001) is the easy plane.
$\theta$ and $\phi$ are defined as the angles from the $c$-axis and the $a$-axis, respectively, see Fig.\ \ref{fig:structure}.
In (b), magnetocrystalline anisotropy within the (001) plane shows that $E$ is minimized near $\phi$=$45^\circ$, which is the $\langle 110 \rangle$ direction. 
The two separate contributions from spin-orbit interactions (SOI) and magnetic dipole-dipole (MDD) interactions are shown in (a), while the in-plane MDD is symmetry-forbidden in (b).
}
\end{figure}
%%%%%%%%%%%%%%%%%%%%%%%%%%%%%%%%%%%%%%%%%%%%%%%%%%%%%%%%%%%%%%%%%%%%%%%%%%%%%%%%%%%%%%%%%

To gain a deeper understanding of this anisotropy, we employed first-principles density functional theory to investigate the magnetocrystalline anisotropy energy (MAE) of Mn$_{2}$Au. 
In Fig.\ \ref{fig:Aniso-DFT}, we present the MAE results for N{\'e}el vector orientations that include $c$ components (out of the (001) plane, which are unfavorable), and orientations within the easy (001) plane.
Energies are plotted as a function of angles $\theta$ and $\phi$ deviating from the $c$-axis and $a$-axis, respectively (see Fig.\ \ref{fig:structure}).
Visually, it is clear that in-plane spin configurations are preferred, while the in-plane angular dependence is comparatively weak. 
The origin of the MAE in Mn$_{2}$Au can be attributed to the spin-orbit interaction (SOI), which plays a more significant role than the magnetic dipole-dipole interaction (MDD).

The MAE coefficients were determined by fitting the calculated data from Fig.\ \ref{fig:Aniso-DFT} to 
\begin{equation}
\label{eq:MAE}
\frac{E_{\text{MAE}}}{V}=K_{1}\sin^2 \theta + K_{2}\sin^4 \theta + K_{22}\sin^4 \theta \cos(4\phi).
\end{equation}
From this we obtain the total in-plane coefficient $K_{1}^{\mathrm{tot}} = 6574$\,kJ/m$^{3} = 1978.3$\,$\mu$eV/f.u. and the total out-of-plane coefficient $K_{22}^{\mathrm{tot}} = 26$\,kJ/m$^{3} = 7.89$\,$\mu$eV/f.u..
These MAE values are one or two orders of magnitude larger than for the metallic antiferromagnets Fe$_{2}$As\cite{yang_magnetocrystalline_2020} and Cr$_{2}$Al\cite{zhao_-plane_2021}.
Both spin-orbit interaction (SOI) and magnetic dipole-dipole (MDD) interaction contribute to the MAE, and their respective contributions to the in-plane and out-of-plane MAE coefficients are as follows:
$K_{1}^{\mathrm{SOI}} = 6144$\,kJ/m$^{3} = 1849.0$\,$\mu$eV/f.u., $K_{1}^{\mathrm{MDD}} = 430$\,kJ/m$^{3} = 129.3$\,$\mu$eV/f.u., $K_{22}^{\mathrm{SOI}} = 26$\,kJ/m$^{3} = 7.89$\,$\mu$eV/f.u., and $K_{22}^{\mathrm{MDD}} = 0$\,kJ/m$^{3}$.
The magnitude of $K_{1}^{\mathrm{MDD}}$ is comparable to that of Fe$_{2}$As\cite{yang_magnetocrystalline_2020} and Cr$_{2}$Al\cite{zhao_-plane_2021}.
However, the absence of $K_{2}^{\mathrm{MDD}}$ is solely due to magnetic symmetry in the (001) plane.
On the other hand, the presence of the heavy element Au in Mn$_{2}$Au significantly enhances the spin-orbit coupling effect, resulting in remarkably large values for both $K_{1}^{\mathrm{SOI}}$ and $K_{22}^{\mathrm{SOI}}$, about 20 and 100 times larger than $K_{1}^{\mathrm{SOI}}$ and $K_{22}^{\mathrm{SOI}}$ of Fe$_{2}$As\cite{yang_magnetocrystalline_2020}.
The positive sign of $K_{1}$ indicates easy-plane magnetism, while the positive sign of $K_{22}$ signifies an easy axis along $\langle 110 \rangle$.
We observe a small energy bump near $\langle 110 \rangle$ in the in-plane MAE. 
This bump appears to be a higher-order feature that persists even when the number of $\mathbf{k}$-points in the calculation is increased, as illustrated in Fig.\ S9.

Using the energy surface method\cite{kang_polar_2022, Kang_mnpt_susceptibility_2023} and a first-principles approach, we further investigated the anisotropic magnetic susceptibility of Mn$_{2}$Au, revealing subtle variations among different orientations.
This method assumes a single-crystalline structure with a single domain, where the Néel vector orientation is aligned perpendicular to the external magnetic field beyond the spin-flop field\cite{kang_polar_2022, Kang_mnpt_susceptibility_2023}.
We believe that this is well justified because with only one symmetrical magnetic site Mn$_2$Au has a much simpler magnetic structure than Fe$_2$As which we studied before \cite{yang_magnetocrystalline_2020}.
Consequently, this approach allows us to determine the anisotropic magnetic susceptibility during the occurrence of a spin-flop transition.
Since Mn$_{2}$Au possesses only one symmetrical magnetic site, we only need to calculate the energy curve while varying net magnetization corresponding to the tilting angle of the magnetic moment on the Mn atom, as illustrated in Fig.\ S11.
The curvature $a$ of these curves is associated with the perpendicular magnetic susceptibility $\chi_{\perp}$, as described by \cite{Kang_mnpt_susceptibility_2023}
\begin{equation}
\label{eq:magsus}
\chi_{\perp}=\frac{\mu_{0}}{2a-\mu_{0}},
\end{equation}
where $\mu_{0}$ is the vacuum permeability.

Our calculations yield values of $\chi^{\mathrm{[100]}}_{\perp}$=$4.27\times10^{-4}$, $\chi^{\mathrm{[110]}}_{\perp}$=$4.11\times10^{-4}$, and $\chi^{\mathrm{[001]}}_{\perp}$=$4.10\times10^{-4}$ for the anisotropic magnetic susceptibility in the regime with a single in-plane magnetic domain (e.g. above a spin flop transition).
The value of $\chi^{\mathrm{[001]}}_{\perp}$ does not change in the single- or multi-domain case (above or below spin-flop, low or high field) for Mn$_2$Au since spins remain normal to the $c$-axis, and the obtained experimental value of  $4.9 \times 10^{-4}$ (Supplemental Figure S7) is in very good agreement.\cite{supplement}
We estimate an error bar due to $\mathbf{k}$-point convergence of about 0.6$\times10^{-4}$ for these values, which is low enough to reliably conclude a very small anisotropy of the susceptibility.
The subtle differences in our simulated values are consistent with a relatively weak influence of magnetocrystalline anisotropy (MCA), which is on the order of kJ to MJ per cubic meter, see Fig.\ \ref{fig:Aniso-DFT}.
The susceptibility is instead dominated by exchange interactions that exhibit a much larger magnitude on the order of GJ per cubic meter (see Fig.\ S10).
Because of this difference in order of magnitude, any anisotropic contribution from MCA is not visible in the susceptibility above the spin-flop field.
As we discuss next, the magnetic field used to measure susceptibility in this work is clearly below the spin-flop transition.
In this regime, the susceptibility can be anisotropic (consistent with Fig.\,\ref{fig:Aniso}) and, in addition, be influenced by domain effects or magnetic defects.

Finally, we note that the out-of-plane and in-plane spin-flop fields can be determined using\cite{Herak:2010}
\begin{eqnarray}
\label{eq:Ansio}
H_{\mathrm{SF}}^{\mathrm{out}} &= \sqrt{\frac{2K_{1}}{\chi_{\perp}-\chi_{\parallel}}}, \\ H_{\mathrm{SF}}^{\mathrm{in}} &= \sqrt{\frac{16K_{22}}{\chi_{\perp}-\chi_{\parallel}}}.
\end{eqnarray}
By using the DFT-calculated anisotropy coefficients and magnetic susceptibility values at 0\,K with an assumption of $\chi_{\parallel}=0$, we obtain the spin-flop fields of $H_{\mathrm{SF}}^{\mathrm{out}}=201$~T and $H_{\mathrm{SF}}^{\mathrm{in}}= 36$~T, respectively.
These results agree with our lack of observation of an in-plane spin flop up to 7~T, and the small amount of domain orientation at 30~T observed in thin films.\cite{sapozhnik_direct_2018}
{\color{black}
The out-of-plane spin flop transition would not be observed experimentally in Mn$_2$Au, since it would always be more stable for the spins to point along the basal plane direction normal to the field. 
}
Higher-field magnetization experiments on bulk Mn$_2$Au could confirm the in-plane prediction.

%%%%%%%%%%%%%%%%%%%%%%%%%%%%%%%%%%%%%%%%%%%%%%%%%%%%%%%%%%%%%%%%%%%%%%%%%%%%%%%%%%%%%%%%%

\section{Conclusions}
In summary, we report the bulk single crystal growth and anisotropic characterization of the metallic antiferromagnet Mn$_2$Au in Bi flux.
Four-point resistivity measurements on bulk single crystals confirm the metallic transport behavior with RRR~=~41 attesting to the low defect concentration in the crystals.
Single crystal neutron diffraction analysis agrees with an easy $ab$-plane, while
the orientation within the plane cannot be determined by low-field susceptibility or neutron scattering measurements due to tetragonal symmetry.
The comparison between experiment and theory for the magnetic arrangement, Mn moment size, and overall susceptibility are consistent.
The computational result indicates an easy axis along $\langle 110 \rangle$, in agreement with prior experiments on thin films.
Rigorous convergence testing and multiple experimental trials confirm both results. 
Further investigation is necessary to determine if there are additional subtle effects that affect the N\'{e}el vector orientation in experimental and computational results, since the anisotropy is small and could be influenced by strain from substrates, surface features, or defects. 
High-field magnetometry and torque magnetometry are additional plausible next steps.
The absence of evidence of an easy-plane spin-flop transition within the range of magnetic measurements up to 7~T is corroborated by computational prediction of the spin-flop field to be 36~T. The high spin-flop fields are likely to be increased by defects and applied strains, and together with the rather isotropic magnetic susceptibility (even along the $c$-axis) pose a unique challenge to fundamental and applied studies that seek to control antiferromagnetic domain structures in predictable ways.

\begin{acknowledgments}
This work was undertaken as part of the Illinois Materials Research Science and Engineering Center, supported by the National Science Foundation MRSEC program under NSF Award No.\ DMR-1720633.
The characterization was carried out in part in the Materials Research Laboratory Central Research Facilities, University of Illinois. 
Single-crystal neutron diffraction was conducted at ORNL's High Flux Isotope Reactor, sponsored by the Scientific User Facilities Division, Office of Basic Energy Sciences, U.S.\ Department of Energy.
This work made use of the Illinois Campus Cluster, a computing resource that is operated by the Illinois Campus Cluster Program (ICCP) in conjunction with the National Center for Supercomputing Applications (NCSA) and which is supported by funds from the University of Illinois at Urbana-Champaign.
This research is part of the Blue Waters sustained-petascale computing project, which is supported by the National Science Foundation (awards OCI-0725070 and ACI-1238993) and the state of Illinois.
Blue Waters is a joint effort of the University of Illinois at Urbana-Champaign and its National Center for Supercomputing Applications.
\end{acknowledgments}

\bibliography{Mn2Au}

\end{document}